\documentclass[article]{elsarticle} 

\usepackage{natbib}
\usepackage{cleveref}
\usepackage{amssymb}
\usepackage [latin1]{inputenc}
\journal{Astroparticle Physics}

\begin{document}

\begin{frontmatter}

\title{The low rate of Galactic pevatrons}

\author{Pierre Cristofari\corref{pierre.cristo}, Pasquale Blasi}
\address{Gran Sasso Science Institute, via F. Crispi 7--67100, L'Aquila, Italy \\
 INFN/Laboratori Nazionali del Gran Sasso, via G. Acitelli 22, Assergi (AQ), Italy}

\author{Elena Amato}
\address{
INAF-Osservatorio Astrofisico di Arcetri, Largo E. Fermi 5, 50125 Firenze, Italy \\
Dipartimento di Fisica e Astronomia, Universita degli Studi di Firenze,
Via Sansone 1, 50019 Sesto Fiorentino (FI), Italy
}




\begin{abstract}
Although supernova remnants remain the main suspects as sources of Galactic cosmic rays up to the knee, the supernova paradigm still has many loose ends. The weakest point in this construction is the possibility that individual supernova remnants can accelerate particles to the rigidity of the knee, $\sim 10^{6}$ GV. This scenario heavily relies upon the possibility to excite current driven non-resonant hybrid modes while the shock is still at the beginning of the Sedov phase. These modes can enhance the rate of particle scattering thereby leading to potentially very--high maximum energies. Here we calculate the spectrum of particles released into the interstellar medium from the remnants of different types of supernovae. We find that only the remnants of very powerful, rare core--collapse supernova explosions can accelerate light elements such as hydrogen and helium nuclei, to the knee rigidity, and that the local spectrum of cosmic rays directly constrains the rate of such events, if they are also source of PeV cosmic rays.  
On the other hand, for remnants of typical core--collapse supernova explosions, the Sedov phase is reached at late times, when the maximum energy is too low and the spectrum at very--high energies is very steep, being mostly produced during the ejecta dominated phase. For typical thermonuclear explosions, resulting in type Ia supernovae, we confirm previous findings that these objects can only produce cosmic rays up to $ \lesssim10^{5}$ GeV. The implications for the overall cosmic ray spectrum observed at the Earth and for the detection of PeVatrons by future gamma--ray observatories are discussed. 
\end{abstract}

\begin{keyword}
\texttt{cosmic ray acceleration - supernova remnants - Galactic}
\MSC[2010] 00-01\sep  99-00
\end{keyword}

\end{frontmatter}


\section{Introduction} The role of supernova remnants (SNRs) for the origin of Galactic cosmic rays (CRs) is considered so well established that it has been elevated to the rank of a paradigm. Despite the solid evidence that particle acceleration takes place at SNR shocks, the paradigm remains based on some poorly known assumptions and certainly fails to explain some observational facts. 

The main challenges that the SNR-CR paradigm faces can be summarised as follows: 1) in most cases in which there is evidence for gamma--ray emission of hadronic origin the spectrum of accelerated particles is steeper than the canonical $E^{-2}$ spectrum predicted by the test particle theory of diffusive shock acceleration (DSA), and much steeper than predicted based on the non-linear generalization of the theory; 2) steeper spectra of CRs injected by SNRs are also required based on the observed B/C ratio and other indicators \cite{evoli2019}; 3) the possibility for SNRs to accelerate high energy CRs relies on the fact that a fast growing CR induced instability may be excited and leads to an enhanced scattering rate, resulting in turn into a higher maximum energy. Even in the presence of this fast growing instability, several factors limit the possibility to explain the observed fluxes in the PeV region, as discussed below. 

The third is certainly the most severe problem that the paradigm encounters and, to say the least, raises the important question of which SNRs, if any, can actually accelerate particles to PeV energies \cite{sveshnikova2003,2019IJMPD..2830022G,martina,ptuskin2010,2013MNRAS.431..415B}. 

{  The answer to this question is instrumental to constructing a solid, credible SNR paradigm for the origin of CRs, or otherwise look for alternative accelerators capable of reaching the PeV range, such as  e.g.  sources in the Galactic Center region, star clusters, or OB associations~(see e.g.~\cite[][]{jouvin2017,aharonian2019,bykov2020}). The problem of the production of PeV particles is one of the key scientific cases of upcoming gamma--ray observatories, such as the Cherenkov Telescope Array (CTA). 
}

Here we consider the three most interesting types of supernova (SN) explosions from the point of view of CR acceleration: typical thermonuclear SNe, typical core--collapse SNe, and peculiar highly energetic core--collapse SNe. We refer to these sources respectively as type Ia, type II, and type II$^{*}$.  We investigate the maximum energy and the spectrum of particles liberated into the ISM both in the form of escape flux (at any given time) and particles accumulated downstream of the shock for the whole duration of the expansion and eventually injected into the ISM, at the end of the SNR life. In the latter case, the effect of adiabatic energy losses is taken into account. For core--collapse SNe the environment is determined by the wind (or winds) of the progenitor star and such environment severely affects the maximum energy and the spectrum of the injected CRs. 

{  Similar explorations of the potential of different types of SNRs to accelerate high energy CRs have been performed before, e.g. by \cite{sveshnikova2003} who tried to identify the most likely class of SNe acting as PeVatrons based on a phenomenological prescription for the maximum energy, previously put forward by \cite{Ellison}. Such work preceded the seminal work by \cite{bell2004} on the growth of the non-resonant CR induced streaming instability, that has profoundly changed this field of investigation and that is at the basis of the present work In Ref. \cite{ptuskin2005,ptuskin2010} different classes of SNRs were also investigated as sources of PeV CRs. The estimates of the maximum energy were based upon the comparison of the acceleration time (or diffusion length, with Bohm diffusion assumed) and the SNR age (or size). As we discuss below these assumptions lead to a large overestimate of the maximum energy. Finally, Refs. \cite{2013MNRAS.431..415B,2013MNRAS.435.1174S,2014MNRAS.437.2802S} also considered core collapse and type Ia SNRs in order to infer the spectrum of CRs contributed by these sources. However the complex structure of the environment in which core collapse SN explosions take place was not taken into account and only CRs escaping the remnant from upstream were included, while we show that CRs advected downstream and losing energy adiabatically have a considerable impact on the shape of the CR spectrum injected into the ISM}.

We find that only very powerful SN explosions (what we call type II$^{*}$ SNe) occurring while the progenitor star  is in the red supergiant (RSG) phase can account for a maximum energy at the beginning of the Sedov phase that is in the PeV range. This happens only for a total explosion energy  $E_{\rm SN}\sim (5\div10)\times 10^{51}$ erg, a mass--loss rate of the progenitor star $\dot{M}\sim 10^{-4}M_{\odot}~\rm yr^{-1}$, an ejecta mass $M_{\rm ej}\sim1$ M$_{\odot}$, and assuming an acceleration efficiency $\xi \sim 10\%$. Interestingly, the rate of SNe of this type that is required to fit the proton flux in the PeV region is of order one every $10^{4}$ years, compatible with the expected rate of rare very energetic events. 

For remnants of ordinary core--collapse SNe, the parameters are much less extreme and we find that the spectrum of particles with very--high energies injected into the ISM is extremely steep and unlikely to contribute to the CR flux at energies above $\sim 1$ TeV. For type Ia SN explosions, the environment is typically rather homogenous and we confirm previous findings, suggesting a maximum energy below $\sim 100$ TeV. 

{  The article is organised as follows: in \S \ref{sec:cr} we describe the basic physics of particle acceleration and escape from SNRs, and the environment expected around different types of SN progenitors, which in turn affects the maximum achievable energy. In \S \ref{sec:transport} we discuss the model used to describe the transport of CRs in the Galaxy. Our results are illustrated in \S \ref{sec:results}. We conclude in \S \ref{sec:conclude}.}

\section{CRs from SNRs} 
\label{sec:cr}

We use the standard test-particle theory of DSA to describe particle acceleration at a SNR shock. The spectrum of particles accelerated at a given time $t$ is $f_0(p,t) \propto (p/p_{\rm inj})^{-\alpha} \exp(-p/p_{\rm max}(t))$, with a normalisation constant calculated so that the energy density at the shock in the form of CRs is a fraction $\xi$ of the ram pressure $\rho u_{\rm sh}^{2}$, where $\rho$ is the density of the gas in which the shock is propagating at speed $u_{\rm sh}$. The acceleration history of the SNR is integrated from early times in the ejecta dominated (ED) phase to the end of the Sedov-Taylor (ST) phase, while we assume that no acceleration takes place during the radiative phase, because of the low shock speed at that time and because substantial ion-neutral damping of waves is expected. 

The spectrum of CRs that a SNR injects into the ISM is the sum of two contributions: 1) the escape flux, made of particles released at any given time from the upstream region in a narrow range of momenta around $p_{\rm max}(t)$, and 2) the particles that are collected downstream of the shock and released into the ISM at the end of the ST phase, after substantial adiabatic energy losses. Below we discuss the two contributions separately. 

We follow the approach of Ref. \cite{caprioli2009} to describe the escape of particles from upstream: the escape flux at time $t$ is concentrated around the maximum momentum, $p_{\rm max}(t)$ reached at that time. The spectrum of escaped particles integrated over the expansion history (starting from a time $t_{0}$ early enough in the ED phase) is:
\begin{equation}
\label{eq:Nesc}
N_{\rm esc}(p) =  \int_{t_0}^{T_{\rm SN}} \;   dt \frac{4 \pi}{\sigma} r^2_{\rm sh}(t) u_{\rm sh}(t)  f_0(p,t) G(p,p_{\rm max}(t)),
\end{equation}
where $\sigma=4$ is the compression factor at the shock, and $G(p,p_{\rm max}(t)) = \exp(-p_{\rm max}(t)/p)/(1-  \exp(-p_{\rm max}(t)/p) )$ is derived in test-particle theory of DSA, although can be easily generalized to its non--linear extension \cite{caprioli2009}. It has been discussed in much previous literature that the escape flux integrated in the ST phase is $\sim p^{-2}$, while at momenta $p\gtrsim p_{M}$, where $p_{M}$ is the maximum momentum reached at the beginning of the ST phase, it steepens considerably, by an amount that depends on how $u_{\rm sh}$ evolves in time \cite{martina,2013MNRAS.431..415B}. 

Particles accelerated to $p<p_{\rm max}(t)$ at any given time are advected downstream and lose energy adiabatically due to the expansion. These particles are only liberated at time $t=T_{SN}$, that we assume coincides with the end of the ST phase, and their spectrum integrated over time is:
\begin{equation}
\label{eq:Nacc}
N_{\rm acc}(p) dp =  \int_{t_0}^{T_{\rm SN}} \;   dt  \frac{4 \pi}{\sigma}  r^2_{\rm sh}(t) u_{\rm sh}(t)  f_0(p',t) dp', 
\end{equation}
where the momentum $p'$ is related to $p$ through the rate of adiabatic losses \cite{caprioli2011}:
\begin{equation}
\label{eq:loss}
\frac{dp}{dt}=  \frac{p}{{\cal L}(T_{\rm SN}, t) } \frac{d{\cal L}(T_{\rm SN}, t) }{dt},
\end{equation}
 and ${\cal L}(T_{\rm SN},t) = (\rho(t) u_{\rm sh}^2(t)/ \rho(T_{\rm SN}) u_{\rm sh}^2(T_{\rm SN}) ) )^{1/3\gamma}$. 

\begin{figure}[b]
\includegraphics[width=.8\textwidth]{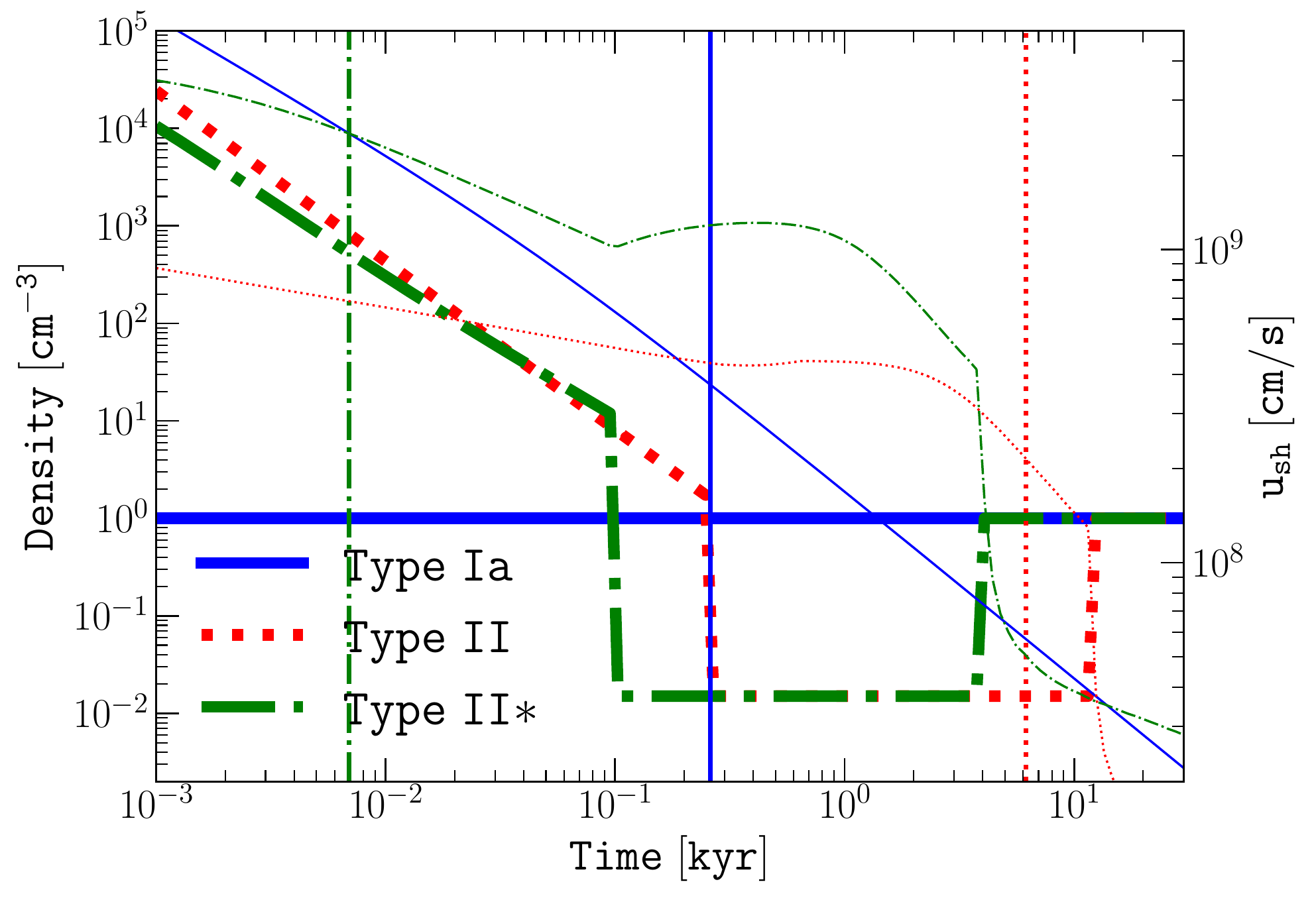}
\caption{\label{fig:rho} Density upstream of the expanding SNR shock (thick) and shock velocity (thin) as a function of time, for type Ia (solid blue), II (dotted red) and II$^{*}$ (dot--dashed green) progenitors of Tab.~\ref{tab:parameters}, assuming $\xi=0.1$. The vertical lines indicate the beginning of the ST phase for each case.}
\end{figure}

We consider shocks resulting from the explosion of thermonuclear (type Ia) and core--collapse supernovae (labelled as type II here). {  Type Ia SN explosions typically occur in a rather uniform ISM, whereas type II SNe evolve} in a structured medium which arises from the final stages of the evolution of the progenitor star~\cite{dwarkadas2005}. During the main sequence stage, the progenitor star produces a wind that excavates a bubble of low density hot gas in pressure balance with the ISM. When reaching the RSG stage, a low--velocity dense wind forms. When the SN explosion occurs, the shock propagates first through a RSG wind with density $n_{\rm w}= \dot{M}/(4 \pi m u_{\rm w} r^2)$, then in a low density hot bubble, and finally reaches the unperturbed ISM. As discussed in \cite{dwarkadas2005}, it may happen that the SN explosion occurs when the progenitor is in the WR phase and the wind is much faster and more diluted than that of RSG stars. The SN explosions that occur in this case typically lead to very steep spectra because they reach the ST phase very late, when the maximum energy is  low.  For a typical core--collapse progenitor, the wind mass--loss rate is $\dot{M}\approx 10^{-5}$M$_{\odot}$yr$^{-1}$, the wind speed is $u_{\rm w}\approx 10^6$ cm s$^{-1}$, the bubble density is $n_{\rm b}=10^{-2}$~cm$^{-3}$, the temperature is $T_{\rm b}=10^{6}$~K. The boundary $r_1$ between the RSG wind  of density $n_{\rm w}= \dot{M}/(4 \pi m u_{\rm w} r^2)$ and the low density bubble is set by equating the RSG wind pressure to the thermal pressure of the  hot bubble, so that $r_1=\sqrt{ \dot{M} u_{\rm w}/4 \pi k n_{\rm b} T_{\rm b}}$,  where $k_B$ is Boltzmann's constant and $m=m_p (1+4 f_{\rm He})/(1+f_{\rm He})$ is the mean mass of interstellar nuclei per hydrogen nucleus, and we adopt a Helium abundance $f_{\rm He}=10\%$.  The boundary between the low density bubble and the ISM depends on the properties of the main sequence star wind~\cite{weaver1977} and is typically $r_2\approx 30$~pc.  The density profile in which the SNR shock expands, and the velocity of the shock are shown in Fig.~\ref{fig:rho} for the three {  benchmark cases considered below}. 

The temporal evolution of $r_{\rm sh}$ and $u_{\rm sh}$ is described by self--similar solutions, in the case of type Ia and type II in the RSG wind~\cite{chevalier1982,tang2017}. In the structured medium, the thin--shell approximation is adopted to derive semi--analytical descriptions~\cite{ostriker1988,BG1995,ptuskin2005}. As we discuss below, for rare core-collapse SNe, that we denoted generically as type II$^{*}$, the parameters of the wind and bubble can be quite different from the typical ones listed above. 

{  In Fig.~\ref{fig:rho} the thick (thin) lines show the density (shock velocity) at the location reached by the shock at time $t$. The fact that the radius does not scale linearly with time, even during the ED phase, explains why the curves for the density in the wind for type II SNRs have a slightly different slope (they would be $\propto r^{-2}$ in the wind of the progenitor star, if plotted as functions of the distance from the explosion center). For core collapse SNe, about a few hundred years after the explosion, the shock enters a hot dilute bubble, in which the shock speed remains roughly constant. Finally, a few thousand years after the explosion, the shock reaches the ISM and its speed drops considerably. Typically, for ordinary type II SNe the ST phase is reached when the shock propagates in the bubble. For very powerful, type II* SNe, the shock velocity is very fast to start with, the wind density close to the star is large and the ST phase is reached very early, typically within a few years from the time of explosion. These considerations are of pivotal importance in the determination of the time dependence of the maximum momentum of accelerated particles, $p_{\rm max}$.
}

\begin{figure}[b]
\includegraphics[width=.8\textwidth]{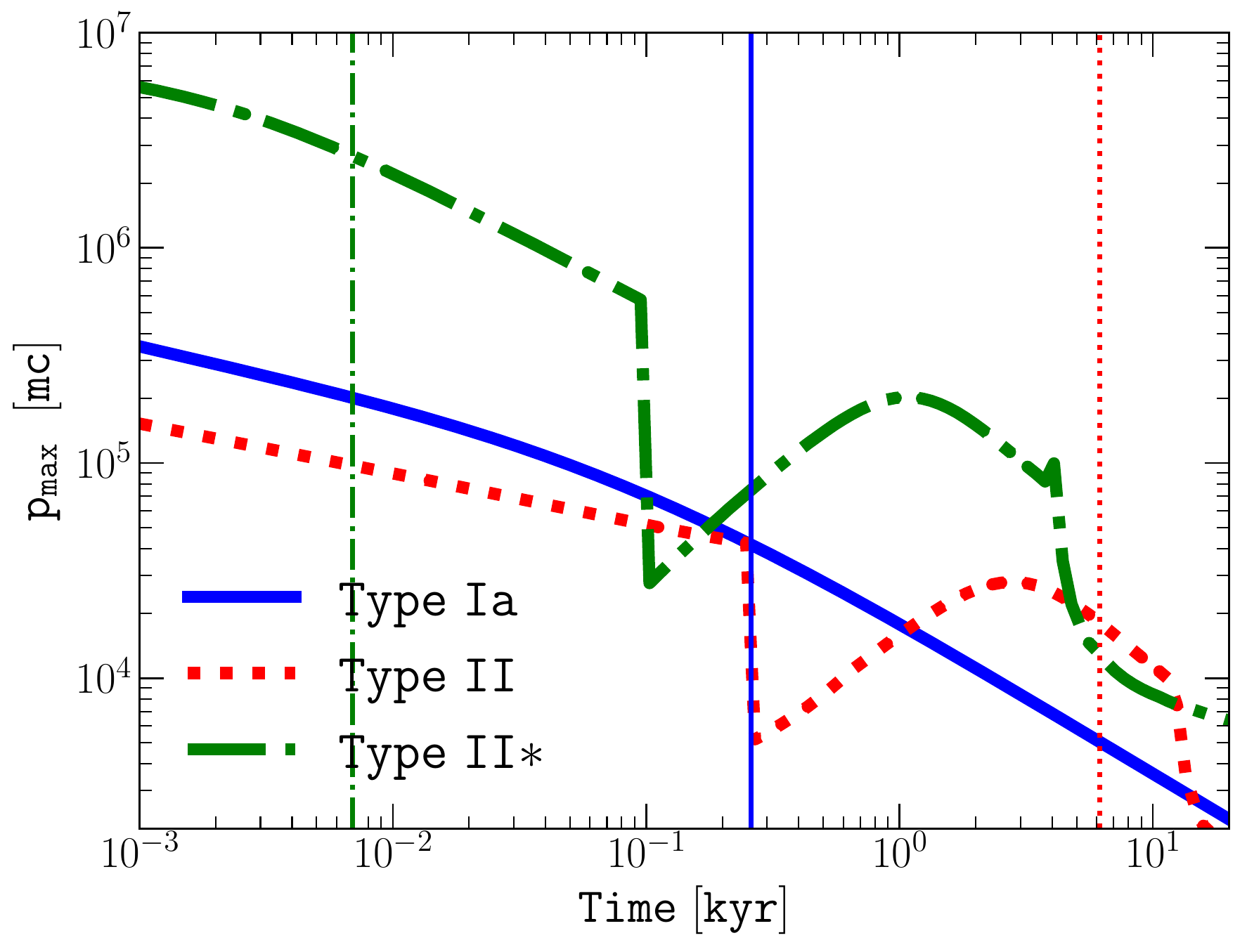}
\caption{\label{fig:emax} Time evolution of the maximum momentum of accelerated protons for type Ia (solid blue), II (dotted red) and II$^{*}$ (dot--dashed green) progenitors of Tab.~\ref{tab:parameters}, assuming $\xi=0.1$. The vertical lines indicate the beginning of the ST phase for each case.}
\end{figure}

\begin{figure}[b]
\includegraphics[width=.8\textwidth]{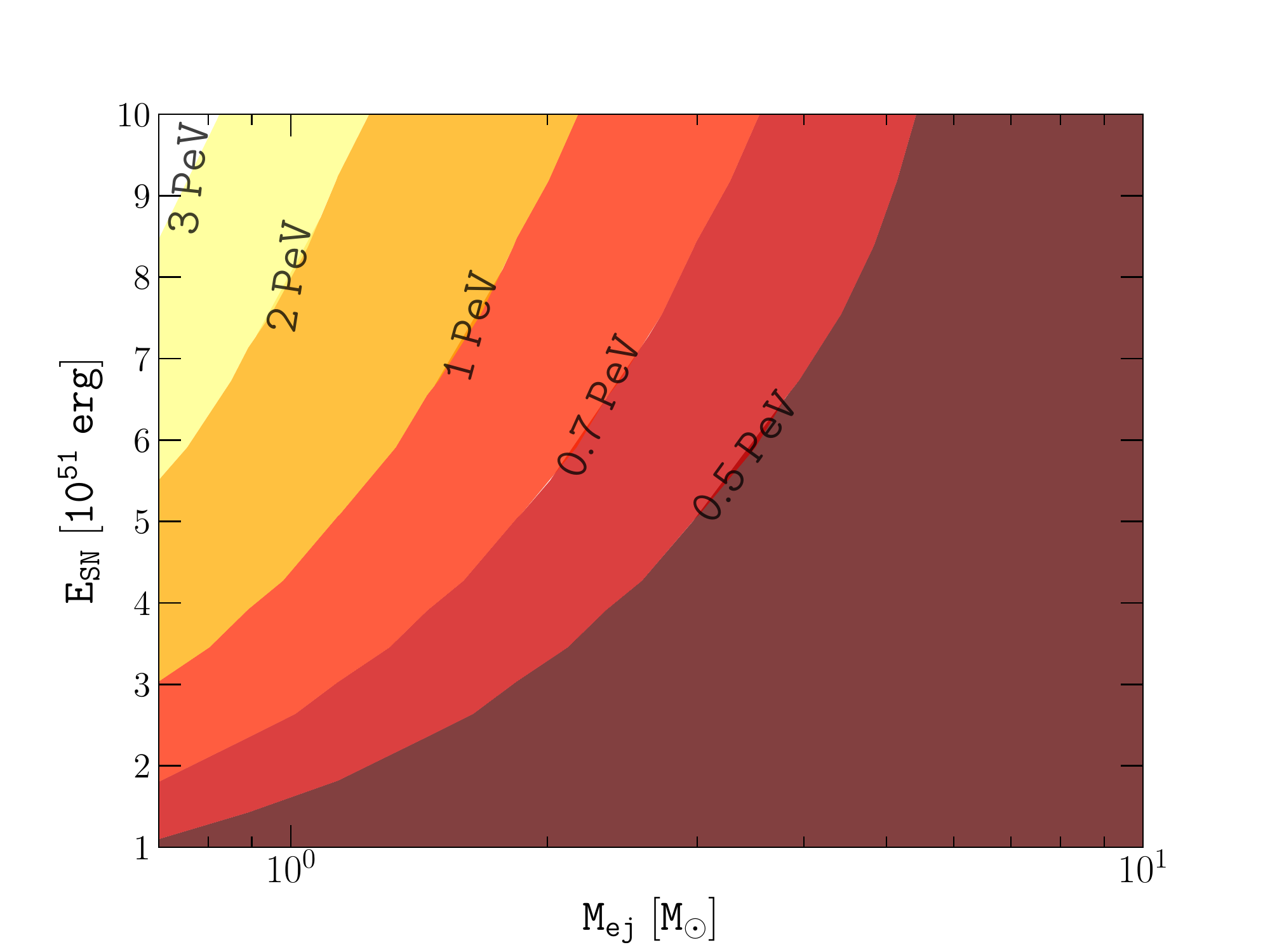}
\caption{\label{fig:contour} Maximum energy of accelerated protons at the transition between the ED and ST phase of a SNR from a type II$^{*}$ progenitor, for different SN total explosion energy $E_{\rm SN}$ and ejecta mass $M_{\rm ej}$, for a RSG mass--loss rate $\dot{M}=10^{-4}~M_{\odot}~\rm yr^{-1}$ and $\xi=0.1$.}
\end{figure}

In the absence of wave self-generation, induced through excitation of a streaming instability, the values of $p_{\rm max}$ that can be achieved are too low to have an impact on the origin of CRs \cite{lagage1,lagage2}. The fastest growing modes excited by CR streaming ahead of a shock are the non-resonant hybrid modes discussed first in Ref. \cite{bell2004}. Their non resonant nature is best expressed by the fact that the wavenumber where the growth rate is the highest, $k_{\rm max}=\frac{4\pi}{cB_{0}}J_{\rm CR}(>p)$, is required to be much larger than $1/r_{\rm L}(p)$, where $r_{\rm L}$ is the Larmor radius  in the unperturbed field $B_0$, of the particles with momentum $p$ dominating the CR current, $J_{\rm CR}(>p)$. Provided this condition is fulfilled, the growth rate is $\gamma_{\rm max}=k_{\rm max} v_{A}$, where $v_{A}=B_{0}/(4\pi\rho)^{1/2}$ is the Alfv\'en speed. As discussed elsewhere {  (see \cite{ABreview,2013A&ARv..21...70B,2014IJMPD..2330013A,NCim} for recent reviews)}, the growth of the instability is accompanied by a growth of the size of the eddies and the saturation occurs when $k_{\rm max}\approx 1/r_{L,\delta B}$, namely when the Larmor radius in the amplified field $\delta B$ becomes equal to $k_{\rm max}^{-1}$. The particles are then able to scatter off these perturbations. Typically the saturation corresponds to a few e-folds, say 5, of the growth:
\begin{equation}
\int_{0}^{t} dt' \gamma_{\rm max}(t') \simeq 5. 
\label{eq:condition}
\end{equation}
As discussed by \cite{2013MNRAS.431..415B}, this is an integral equation for $p_{\rm max}(t)$ that can be solved analytically, and results in: 
\begin{equation}
\label{eq:pmax}
p_{\rm max}(t) \approx \frac{r_{\rm sh}(t)}{10} \frac{\xi e \sqrt{4 \pi \rho(t)} }{\Lambda} \left(\frac{u_{\rm sh}(t)}{c}\right)^{2},
\end{equation}
where $\Lambda=\ln\left(\frac{p_{\rm max}(t)}{mc}\right)$. 

 When $\xi \rho(t) u_{\rm sh}^3(t)/(\Lambda c)<B_0^2/(4\pi)$, the non--resonant modes cannot be excited, and we assume that the maximum energy is determined by the growth of resonant modes. For simplicity here we neglect the role of damping processes in limiting the growth of resonant modes. Damping contributes to further lowering the maximum energy during late stages of the SNR evolution.

As we discussed above, the effective maximum momentum, $p_{\rm M}$, that appears in the spectrum of CRs released by the SNR into the ISM, is the one reached at $t$ corresponding to the transition from ED to ST phase. For typical parameters of a type Ia SNR, $p_{\rm M}\simeq 50$ TeV/c, quite below the energy of the knee, as shown in Fig. \ref{fig:emax} (solid curve), that illustrates the time dependence of $p_{\rm max}$. The same figure also shows the maximum energy for the case of typical core--collapse SNRs  (dotted curve), for the reference values of the parameters listed in Tab.~\ref{tab:parameters} and assuming $\xi=0.1$. Even in this case, for typical conditions, it is not feasible to reach the energy of the knee, at any time during the evolution of these SNRs. 
For the type II$^{*}$ SN explosions the conditions may potentially be more promising because of the larger energy of the explosion and the larger expected mass--loss rate for these events. If to assume that in the last $\sim 10^{5}$ years of the RSG phase of the progenitor the mean mass--loss rate is $\sim 10^{-4}M_{\odot}\rm yr^{-1}$, the maximum energy at the beginning of the ST phase as a function of {  ejecta mass and total explosion energy} is shown in Fig.~\ref{fig:contour}, assuming an acceleration efficiency $\xi=0.1$. 

Fig.~\ref{fig:contour} illustrates in a clear way how, even in the presence of these extreme conditions, reaching PeV maximum energy at the beginning of the ST phase, that now takes place only a few tens of years after the explosion in the RSG wind, a very large explosion energy is required ($E_{\rm SN}\sim 4\div 10\times 10^{52}$ erg) and relatively low ejecta mass (at most few solar masses). The thick dash-dotted curve in Fig.~\ref{fig:emax} shows the time dependence of $p_{\rm max}(t)$ for the rather optimistic case of $M_{\rm ej}=1M_{\odot}$ and $E_{\rm SN}=10^{52}$~erg. 

%
\begin{table}
\centering
\begin{tabular}{cccc}
\hline
 Type &  Ia &  II & II* \\
 \hline
  $M_{\rm ej}$ [M$_{\odot }$]  & 1.4 & 5 & 1 \\ 
     E$_{\rm SN}$ [$10^{51}$ erg]  & 1 & 1 & 10 \\ 
 $\dot{M}$ [$10^{-5}$   M$_{\odot}$/yr]  & -- & 1 & 10 \\ 
u$_{\rm w}$ [10$^{6}$ cm/s ] & -- & 1 & 1 \\ 
r$_1$ [pc]  & -- & 1.5 &  1.3 \\ 
\end{tabular}
\caption{Parameters associated to the three considered progenitors~\cite{ptuskin2005,ptuskin2010}.}
\label{tab:parameters}
\end{table}

In all cases of core--collapse SNe, the shock velocity and location in the complex medium around the progenitor star have been calculated using the thin-shell approximation \cite{ostriker1988,BG1995,ptuskin2005}. 

It is worth stressing that much larger maximum energies were previously quoted by \cite{ptuskin2010}, the main reason being that they assumed some phenomenological description for the strength of the magnetic field in the shock region, at odds with the predictions based on the growth of the non-resonant CR driven modes. {  Moreover, the condition used for the maximum energy, based on the size of the diffusion length, is incompatible with the condition in Eq. \ref{eq:condition},   corresponding to a magnetic field strength a factor $\sqrt{c/u_{\rm sh}}$ larger than what Bell's modes can produce~\cite{NCim}.}.

\section{Galactic transport of CRs} 
\label{sec:transport}

{  Although the main purpose of this article is not to provide a best fit to the observed CR spectrum, it is important to realize that while the maximum energy is mainly determined by the CR acceleration efficiency, $\xi$, the flux of CRs observed at the Earth is proportional to the product of $\xi$ and the rate of occurrence of the given type of SN explosions, so that a description of CR transport in the Galaxy is necessary. This will allow us to assess the role of different types of SN explosions as potential PeVatrons.

Here we adopt a simple \textit{weighted slab model}~\cite{jones2001,aloisio2013} for CR transport. The thin Galactic disk of half--thickness $h_{\rm d}$ and radius $R_{\rm d}=15$ kpc is located at $z=0$ and is assumed to be the site where both sources and gas are located. The equation describing  $I(E)$, the flux of protons with kinetic energy, $E$, can be obtained imposing free escape at the halo boundary $H \gg h_d$, $I(E, z=\pm H)=0$, so that:
\begin{equation}
\label{eq:I}
I(E)\left[\frac{1}{X(E)} + \frac{\sigma}{m}\right]+\frac{d}{dE} \left\{ \left[ \left[ \frac{dE}{dx}\right]_{\rm ad + ion}  \right]I(E) \right\} = \frac{p^2 q_0(p)}{\mu v}
\end{equation}
where the rate of adiabatic and ionization losses,  $ \left[ \frac{dE}{dx}\right]_{\rm ad + ion}$ is detailed in~\cite{evoli2019} and~\cite{mannheim1994}. Here $v$ is the particle velocity,  and  $X$ is the grammage, a function of the diffusion coefficient $D$ and advection velocity $u$: 
\begin{equation}
X(E)= \frac{\mu v}{2 u} \left[1- \exp(-\frac{u H}{D(E)}) \right]
\end{equation}
with $\mu= 2 h_{\rm d} m n_0 ( 1+ f_{\rm He})$. From observations $\mu \approx 2.3$~mg/cm$^2$~\cite{ferriere2001}, and we adopt $n_0$= 1 cm$^{-3}$. 
In agreement with recent results~\cite{genolini2017}, the diffusion coefficient is assumed to have the functional form: 
\begin{equation}
D(p) = D_0\frac{v(p)}{c}  \frac{ \left(p/mc\right)^{\delta} } {\left[ 1+ (p/p_{\rm b})^{\Delta \delta/r}  \right]^{r}}.
\end{equation}
Fitting to available AMS--02 data, \cite{evoli2019}~found a best fit with $r=0.1$, $\Delta \delta= 0.2$ and $p_{\rm b}=312$ GeV/c, $D_0=1.1 \times 10^{28}$cm$^{2}$/s and $\delta=0.63$ assuming $H=4$~kpc and adopting a Galactic advection velocity of $u=7$ km/s.
Finally, the injection $q_0$ is assumed to be due to SNRs. Assuming an explosion rate $\nu_{\rm SN}$, we write: 
\begin{equation}
q_0(p)=\frac{ \nu_{\rm SN} }{\pi R_{\rm d}^2} \left[ N_{\rm acc}(p) + N_{\rm esc}(p) \right],
\end{equation}
where we indicated separately the injection into the ISM of trapped particles, $N_{\rm acc}$, and escaping particles $N_{\rm esc}$. 
}

\section{Results} 
\label{sec:results}

We calculated the flux of CRs from type Ia, II and II$^{*}$ SNRs, using as a boundary that the flux of CRs at the Earth from each of them be at most equal to the observed CR proton flux. This is a very important constraint, that translates into an upper limit on $\xi$, for a given SN rate, thereby constraining $p_{\rm max}$. For a given value of $\xi$, the energy efficiency $\xi_{\rm SN}$, corresponding to the fraction of total explosion energy converted into CRs over the lifetime of a SNR can be defined as: 
\begin{equation}
\xi_{\rm SN}= \frac{1}{E_{\rm SN}} \int_{p_{\rm min}}^{p_{\rm max}} dp \;T(p)  4 \pi  p^2 [ N_{\rm acc}(p) + N_{\rm esc}(p)].
\end{equation}

\begin{figure}[b]
\includegraphics[width=.8\textwidth]{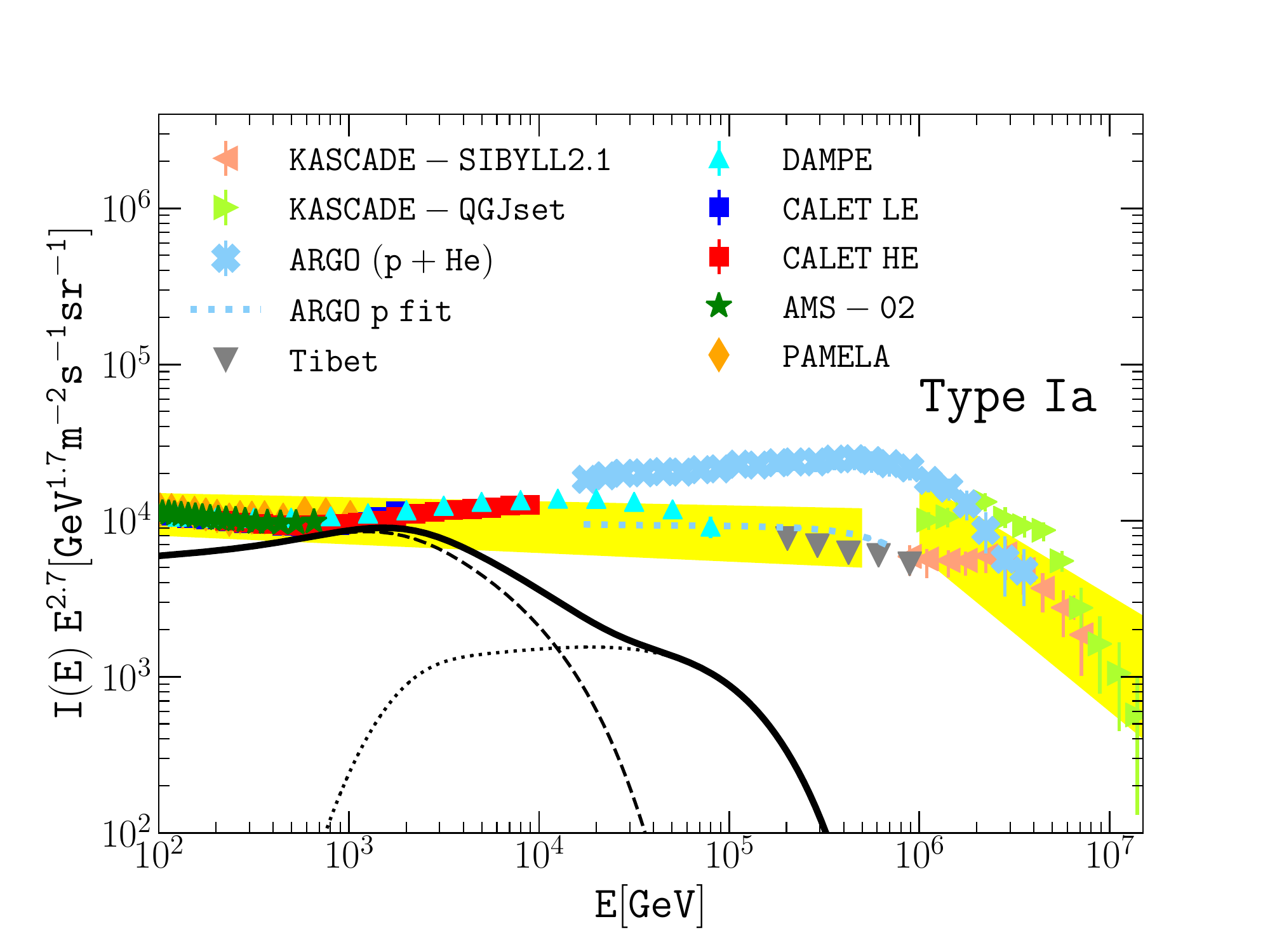}
\caption{\label{fig:typeIa} Galactic CR protons from type Ia SNRs. Contributions from cumulative ac celerated particles $N_{\rm acc}$ (dashed), escaping particles $N_{\rm esc}$ (dotted) and their sum (solid) are shown. $\alpha=4$, $\nu_{\rm SN, Ia}= 1/100$~yr$^{-1}$ and $\xi_{\rm SN}=0.11$ ($\xi=0.10$). Local data from various experiments are shown: AMS-02~\cite{aguilar2015}, PAMELA~\cite{adriani2011}, CALET LE and HE~\cite{CALET2019}, DAMPE~\cite{DAMPE2019}, ARGO--YBJ~\cite{bartoli2015}, ARGO fit for protons~\cite{mascaretti2019}, Tibet~\cite{TIBET2011} and KASCADE~\cite{antoni2005}. The yellow areas correspond to the typical level of measured protons. }
\end{figure}

\begin{figure}[b]
\includegraphics[width=.8\textwidth]{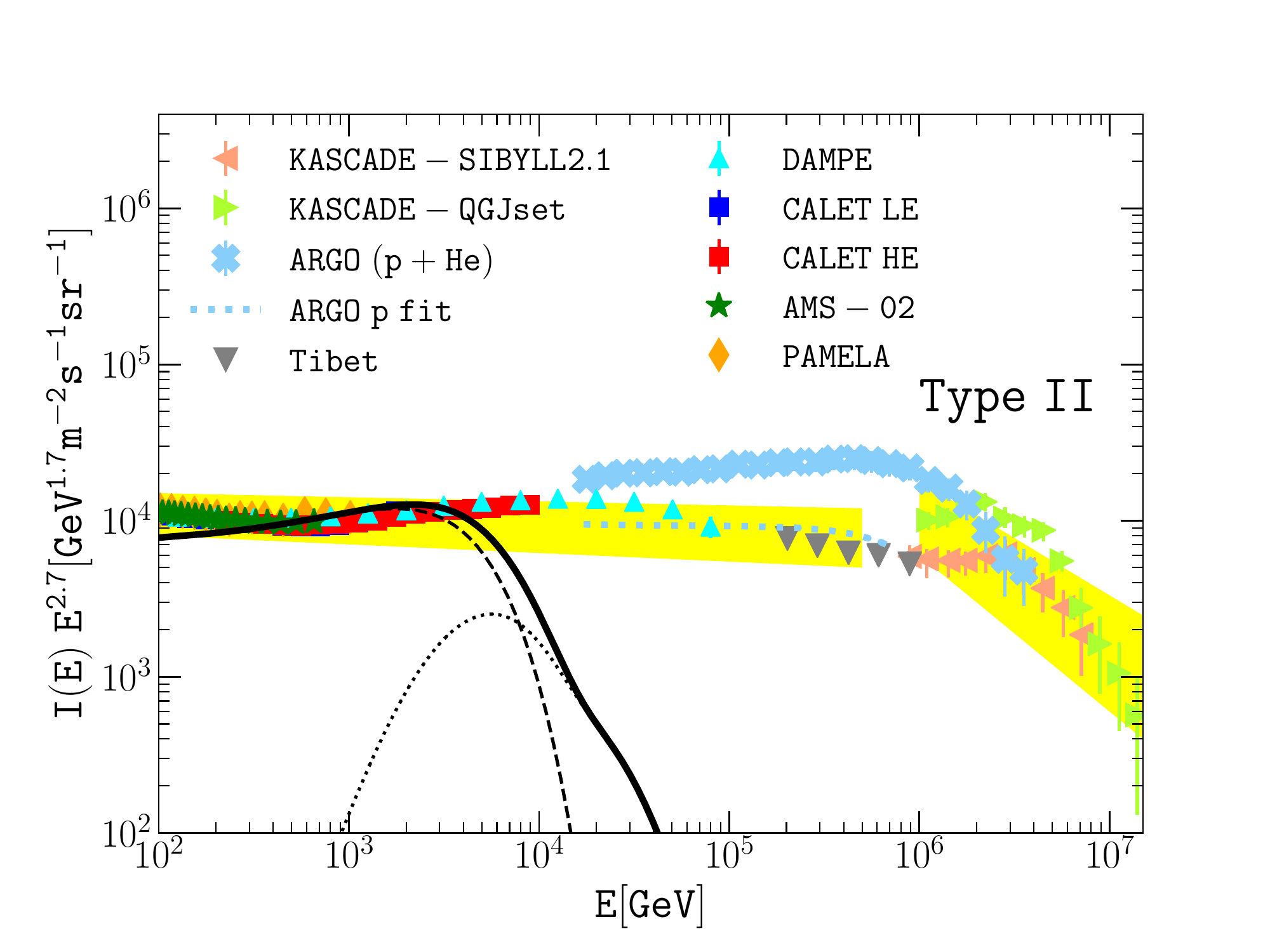}
\caption{\label{fig:typeII} Analog to Fig.~\ref{fig:typeIa} for type II progenitors. $\alpha=4$, $\nu_{\rm SN, II}= 2/100$~yr$^{-1}$ and $\xi_{\rm SN}=0.07$ ($\xi=0.03$). }
\end{figure}

In Fig.~\ref{fig:typeIa} we show our results for type Ia SNRs, for which the well--known rate is $\nu_{\rm SN, Ia}\approx$1/century. The available data on the proton flux (or when not available the flux of light nuclei, H+He) are also shown. The yellow shaded area, shown for illustration purposes, encompasses the range of observed fluxes, with their own systematic uncertainties. The dashed line represents the contribution of CRs trapped in the downstream, after adiabatic losses and transport in the Galaxy, while the dotted line is the contribution of escaped particles. The total flux is the thick solid line. Saturating the CR flux to the observed flux at $\sim 100$ GeV energies requires $\xi_{\rm SN}=0.11$ ($\xi=0.10$), which corresponds to a maximum effective energy in these sources of few tens TeV. The results for the bulk of type II SNRs (rate $\nu_{\rm SN, II}\approx$2/century) are shown in Fig.~\ref{fig:typeII}. The normalisation to the flux in this case requires $\xi_{\rm SN}=0.07$ ($\xi=0.03$). As also visible in Fig.~\ref{fig:emax}, in this case the effective maximum energy is very low, because the ST phase starts late, when the shock propagates in the hot dilute bubble: as a result, {  at very high energies, that here we are most interested in,} the source spectrum is very steep. 

\begin{figure}[b]
\includegraphics[width=.8\textwidth]{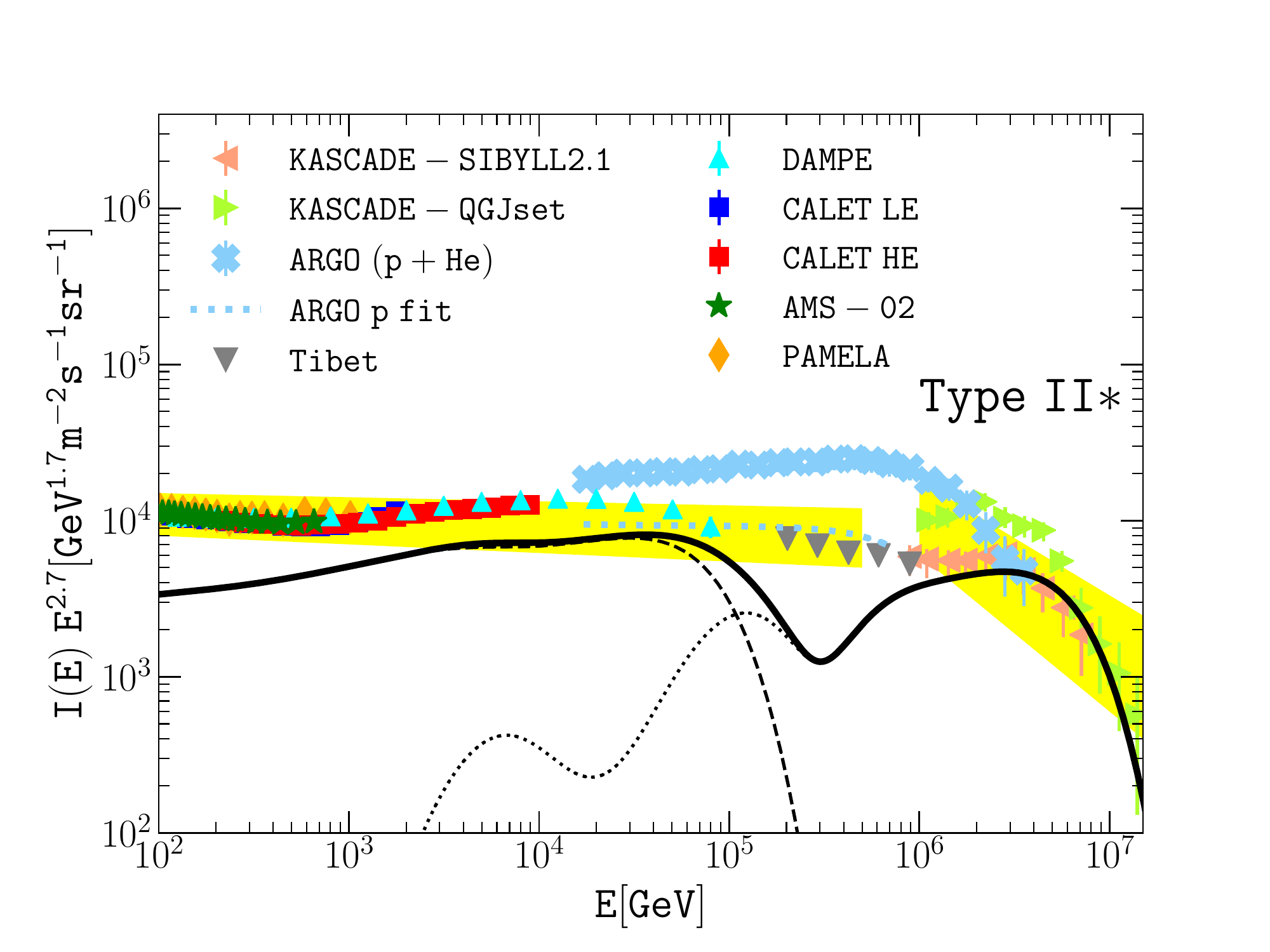}
\caption{\label{fig:typeIIb} Analog to Fig.~\ref{fig:typeIa} for type II* progenitors. $\alpha=4$, $\xi=0.1$ ($\xi_{\rm SN}$=0.22) and $\nu_{\rm SN, II*}= 1.5 \% \times$ 2/100~yr$^{-1}$.}
\end{figure}

The situation is quite different for type II$^{*}$ SN explosions: a maximum energy of 1 PeV is achieved at the beginning of the ST phase if $\xi=0.1$ ($\xi_{\rm SN}=0.22$), a rate of mass--loss ${\dot M} =10^{-4}M_{\odot}~\rm yr^{-1}$, a total explosion energy $E_{\rm SN}=10^{52}$~erg and an ejecta mass of $M_{\rm ej}=1M_{\odot}$ which implies that the observed flux of CR protons is reproduced for a rate $\nu_{\rm SN, II^{*}} \lesssim (1\div 2.5)\% \times 3/100$~yr$^{-1}$ (Fig. \ref{fig:typeIIb}). A more realistic value for $M_{\rm ej}$ of few solar masses would require an even larger energetics, as shown in Fig. \ref{fig:contour}, or { mass--loss rates $\gtrsim 10^{-4} M_{\odot}$/year}.
If these rare SNe are to {  account for} particle acceleration up to the knee energy, they may well contribute most of the CR flux also in the energy region $10^{2}-10^{4}$ GeV, leaving little room for other types of SNe as sources of Galactic CRs. Remarkably, the inferred rate of these SN explosions as derived above is within the typical estimated rate of type IIb and IIn events, $\lesssim$4\% of all Galactic SNe~\cite{smartt2009,leaman2011,dahlen2012,moriya2014,ouchi2017,sravan2019}, with characteristics which are very similar to those of what we named II$^{*}$ but could also correspond to peculiar, rare SNe of other types~\cite{zirakashvili2016}. The transition from CRs trapped in the downstream and those that gradually escape from the SNR at any given time forms a feature at $\sim 10$ TeV that is reminiscent of the DAMPE feature in the proton spectrum \cite{DAMPE2019}. 

\section{Discussion and conclusions} 
\label{sec:conclude}

There are several important conclusions that can be drawn from the calculations illustrated above: 1) if the growth of the non-resonant modes is the main mechanism for magnetic field amplification at SNR shocks, then the maximum energy  is in the PeV range only for rare remnants of very energetic SN explosions, if the efficiency of CR acceleration is $\xi\approx 10\%$.  The requirements in terms of energetics of the SN explosion and mass--loss rate may be made appreciably weaker if PeV CRs are helium nuclei, as can be easily inferred from Fig. \ref{fig:contour}; 2) the rate of these rare events is constrained by the CR proton spectrum, and if their contribution is dominant at the knee, it is also dominant at $10^{2}-10^{4}$ GeV, so that little room is left for more ordinary SNRs at such energies; 3) the spectrum of CRs injected by a given SNR is always the integral over its temporal evolution of two components, the escape flux from upstream and the contribution of CRs advected downstream and eventually liberated into the ISM at the end of the SNR evolution, after substantial adiabatic energy losses. The overlap of the two components leads to the appearance of features in the spectrum that qualitatively resemble the ones recently claimed by DAMPE \cite{DAMPE2019}. 4) The small rate of type II$^{*}$ SNe, about one every $10^{4}$ years, and the fact that PeV energies are typically reached in the first few years after the SN explosion make the perspectives of detecting very--high--energy gamma rays from these potential PeVatrons rather grim and certainly less optimistic than in previous studies \cite{cristofari2018}, at least for observatories such as CTA~\cite{CTAbook}. On the other hand, if CTA were to detect a number of shell SNRs acting as PeVatrons, this may well suggest that other, more effective processes of magnetic field amplification are at work at SNR shocks, which would then reflect in a need to reconsider the very bases of the SNR paradigm for the origin of CRs. 

{  At the same time, while the relatively small rate of potential PeVatrons would suggest to consider constraints coming from anisotropy, it is in fact unlikely that future measurements of the anisotropy in the PeV range will carry useful information about the nature of sources of PeV CRs. This is because the relative rarity of sources will mostly increase the fluctuations rather than the mean anisotropy (see e.g.~\cite[][]{blasi2012}). }

It is also worth noticing that none of the types of SNRs considered here is able alone to describe the relatively smooth CR spectrum that we measure over many decades in energy. In a way, rather than being surprised by the appearance of features, one should be surprised by the fact that the CR spectrum is so regular.  On the other hand, it is not clear how this conclusion may change by taking into account a spread in the values of the explosion energy and rate of mass--loss for each of the SN types considered above. The impact of the points above calls for an urge for questioning some of the basic assumptions. In this sense, one should be aware that in the dense environment of RSG winds, it is unlikely that particles can really escape freely, being their upstream gyro-radius always smaller than the size of the shock for the energies considered here. Perhaps this may imply that some additional acceleration may be taking place even after the non-resonant modes  stop growing. The detection of some of these astrophysical sources in the PeV range by current and future instruments~\cite{HAWC,SWGO} might provide precious clues in this respect.

\section*{Acknowledgments} 
We thank R. Aloisio and C. Evoli for fruitful discussions.  EA acknowledges support from ASI/INAF n. 2017-14-H.O, SKA-CTA-INAF 2016 and INAF-Mainstream 2018.

\bibliography{SNR.bib}

\end{document}